\documentclass[aps,prd,twocolumn,amsmath,amssymb,amsfonts,nofootinbib,superscriptaddress]{revtex4-1}

\usepackage{graphicx}
\usepackage{dcolumn}
\usepackage{bm}
\usepackage[scientific-notation=true]{siunitx}
\usepackage[dvipsnames]{xcolor}
\usepackage{lineno}
\usepackage{xspace}
\usepackage[authormarkup=none]{changes}
\usepackage{comment}

\newcommand{\bea}{\begin{eqnarray}}
\newcommand{\eea}{\end{eqnarray}}
\newcommand{\be}{\begin{equation}}
\newcommand{\ee}{\end{equation}}

\definechangesauthor[color=Purple]{NA}

\newcommand{\TFFT}{T_\text{FFT}}

\newcommand{\Tobs}{T_\text{obs}}

\def\gwh{gravitational-wave\xspace}
\def\gws{gravitational waves\xspace}

\def\cw{continuous wave\xspace}
\def\cwh{continuous-wave\xspace}

\def\cws{continuous waves\xspace}

\begin{document}

\title{Constraints on planetary and asteroid-mass primordial black holes \\ from continuous gravitational-wave searches}

\author{Andrew L. Miller}
\email{andrew.miller@uclouvain.be}
\affiliation{Université catholique de Louvain, B-1348 Louvain-la-Neuve, Belgium}
\author{Nancy Aggarwal}
\email{nancy.aggarwal@northwestern.edu}
\affiliation{Center for Fundamental Physics, Department of Physics and Astronomy,
Northwestern University, Evanston, Illinois 60208, USA}
\affiliation{Center for Interdisciplinary Exploration and Research in Astrophysics (CIERA),
Department of Physics and Astronomy, Northwestern University, Evanston, Illinois 60208, USA}

\author{Sébastien Clesse}
\email{sebastien.clesse@ulb.be}
\affiliation{Service de Physique Th\'eorique, Universit\'e Libre de Bruxelles, Boulevard du Triomphe CP225, B-1050 Brussels, Belgium}
\author{Federico De Lillo}
\email{federico.delillo@uclouvain.be}
\affiliation{Université catholique de Louvain, B-1348 Louvain-la-Neuve, Belgium}

\date{\today}

\begin{abstract}
We present new constraints on the merging rates of planetary-mass and asteroid-mass primordial black hole binaries using limits on \cws (quasi-monochromatic, quasi-infinite duration signals) derived from an all-sky search for isolated compact objects in the first six months of the third observing run (O3a) of LIGO/Virgo. We calculate the merging rates of these binaries in a model-independent way, and convert them to constraints on the primordial black hole abundance with minimal modelling assumptions. Our results show that all-sky searches are sensitive to sources at most {$\mathcal{O}(10$ pc)} away for systems with chirp masses of {$\mathcal{O}(10^{-5}M_\odot)$} at \gwh frequencies around {30-40 Hz}. 
These results also show that \cwh searches could in the future directly probe the existence of planetary-mass and asteroid-mass  primordial black holes, especially those in binaries with asymmetric mass ratios. {Furthermore, they demonstrate that new methods accounting for the full nonlinear gravitational-wave frequency evolution are needed to improve constraints on primordial black holes.}
\end{abstract}

\pacs{Valid PACS appear here}
\maketitle


\section{Introduction}

{The recent detections of \gws by Advanced LIGO \cite{2015CQGra..32g4001L} and Virgo \cite{2015CQGra..32b4001A} have allowed new and unique observations of black holes~\cite{LIGOScientific:2018mvr,LIGOScientific:2020ibl,LIGOScientific:2021usb}.  But the chirpy \gwh signals from black hole or neutron star mergers are likely not the only detectable form of \gws.  Ground-based interferometers can also be used to probe other types of signals, such as \gwh bursts from violent astrophysical phenomena and stochastic \gwh backgrounds,  e.g. those coming from early Universe phenomena at much higher energies than those accessible to particle accelerators.
Continuous waves, defined as quasi-monochromatic and quasi-infinite duration} signals, are among the signals that have so far evaded detection.  They can be emitted by asymmetrically rotating neutron stars, slowly spinning down due to a quadrupolar deformation on their surfaces \cite{horowitz2009breaking,osborne2019gravitational}, or slowly spinning up due to accretion \cite{watts2008detecting}.  {Another possible source of \cws in the frequency range probed by LIGO/Virgo are exotic, planetary-mass compact binaries during their inspiral phase, such as primordial black holes (PBHs)~\cite{Miller:2020vsl,Pujolas:2021yaw}.  Observing PBHs would have groundbreaking implications for cosmology since they could be connected to the still mysterious dark matter, explain some LIGO/Virgo mergers~\cite{Bird:2016dcv,Clesse:2016vqa,Sasaki:2016jop,Carr:2019kxo,Hutsi:2020sol,Clesse:2020ghq,Jedamzik:2020ypm,Jedamzik:2020omx,Hall:2020daa,Franciolini:2021tla,Garcia-Bellido:2020pwq,Chapline:2018sbr}, provide the seeds of super-massive black holes, and shed a new light on the physics at play in the early Universe. Gravitational waves can probe quite a wide mass regime for PBHs, and have already set strong constraints on the fraction of dark matter that PBHs could compose, summarized in \cite{Carr:2020gox,Carr:2020xqk}. In this paper, we use for the first time the upper limits from a \cwh search from the first six months of LIGO/Virgo's third observing run (O3a) to set constraints on the merging rates and abundance of planetary-mass and asteroid-mass PBHs. {In particular, asteroid-mass PBHs are poorly constrained, since previous limits based on femtolensing of gamma ray bursts, optical microlensing, and destruction of white dwarfs by passing PBHs, have been, at best, weakened, and at worst, removed \cite{Katz:2018zrn, Montero-Camacho:2019jte}. }

Planetary chirp mass systems are well-motivated observationally as well as theoretically. There have been recent detections of star and quasar microlensing events~\cite{Niikura:2019kqi,Hawkins:2020zie,bhatiani2019confirmation} that suggest compact objects or PBHs with masses $ [10^{-6}, 10^{-5}] M_\odot$ could constitute a fraction of dark matter of order $f_{\rm PBH} \sim 0.01$, which is consistent within the unified scenario for PBH formation presented in ~\cite{clesse2018seven}, but greater than expected for floating planets \cite{sivaram2019primordial}.  It has even
been hypothesized that Planet 9
could be a PBH with a mass of $ 10^{-6} M_\odot$ that was captured by the solar system \cite{Scholtz:2019csj}, motivating the development of methods to detect the accretion of small Oort cloud {objects} \cite{Siraj:2020upy}. However, astrophysical uncertainties plague these observations, e.g. due to the clustering properties of
PBHs \cite{Garcia-Bellido:2017xvr,Calcino:2018mwh,Belotsky:2018wph,Carr:2019kxo,Carr:2019kxo,Trashorras:2020mwn,DeLuca:2020jug}, underlining the importance of probing these mass regimes using complementary and independent observational methods that could help to distinguish PBHs from other sources. With this in mind, searches for \gws from sub-solar mass binary black hole systems have already been performed \cite{LIGOScientific:2018glc,abbott2019search,Nitz:2020bdb,horowitz2020search,Nitz:2021vqh,Phukon:2021cus,LIGOScientific:2021job} 
using matched filtering techniques, as well as for nonlinear memory effects \cite{Ebersold:2020zah}, but these works only probe a small portion of possible PBH masses{, between $0.1 M_\odot$ and $1M_\odot$}.  {In the future, binary mergers with lower chirp masses could also be detected using resonant high-frequency \gwh detectors~\cite{Herman:2020wao,Aggarwal:2020olq,Aggarwal:2020umq}}.

Much effort in detecting \cws has {until now} gone into developing methods and running searches that target individual pulsars \cite{abbott2020gravitational,abbott2019narrow}, binary pulsar systems \cite{messenger2015gravitational,leaci2015directed}, specific places in the sky with an expected high concentration of neutron stars \cite{piccinni2020directed,dergachev2019loosely}, post-merger remnants \cite{longpmr,Aasi:2014ksa,abbott2017search}, and electromagnetically silent sources anywhere in the sky \cite{sieniawska2019continuous,Riles:2017evm}. In the last few years, these methods have been generalized to search directly for dark matter \cite{guo2019searching,PhysRevLett.121.061102,grote2019novel,Miller:2020kmv,LIGOScientific:2021odm} and 
ultralight boson clouds around black holes \cite{d2018semicoherent,isi2019directed,sun2019search,palomba2019direct}. 
\textit{Transient} \cwh signals \cite{Sarin:2018vsi,lasky2017braking} have {also}
recently garnered interest after the detection of GW170817 \cite{gw170817FIRST}, and a few methods have been developed to detect them \cite{Sun:2018hmm,banagiri2019search,Oliver:2018dpt,PhysRevD.98.102004,miller2019effective,mytidis2015sensitivity}.  {Using \cwh (transient or not) methods to probe planetary-mass black holes during the inspiraling phase was considered for the first time in Ref~\cite{Miller:2020vsl} with the Hough Transform; however, no actual limit on the merger rate or abundance of such black holes have been calculated thus far.}

Indeed, compact objects that are far enough away from each other emit \gws as their orbital frequencies slowly increase over time. Their orbits are quasi-circular until the innermost stable circular orbit (ISCO) \cite{maggiore2008gravitational}, corresponding to a \gwh frequency $f_{\rm ISCO} \simeq 4400 {\, \rm Hz} \times M_\odot / (m_1 + m_2)$.  
From a data analysis point of view, the \cwh signal coming from a planetary-mass inspiraling black hole binary does not differ from a standard \cwh signal from isolated neutron stars. Therefore, upper limits on the strain amplitude {\(h_0(f)\)} obtained from all-sky \cwh searches \cite{abbott2019all,LIGOScientific:2021tsm} can be mapped to limits on the distance reach as a function of chirp mass for black hole binaries.
{These distance reaches can then be used to constrain merging rates and, after assuming a rate model, the PBH abundance.  We follow these steps in this paper.}

\section{Limits on PBH binary distance}

The \gwh signal emitted due to two objects in a quasi-circular orbit has an amplitude $h_0$ given by:
\begin{eqnarray}
h_0&=&\frac{4}{d}\left(\frac{G \mathcal{M}}{c^2}\right)^{5/3}\left(\frac{\pi f_{\rm gw}}{c}\right)^{2/3}~ \label{hamp}\\
&\simeq&1.61\times 10^{-26} \left(\frac{d}{10 \text{ pc}}\right)^{-1} \left(\frac{\mathcal{M}}{10^{-6}M_\odot}\right)^{5/3} \left(\frac{f_{\rm gw}}{50 \text{ Hz}}\right)^{2/3}, \nonumber
\end{eqnarray}
where $d$ is the distance from the source to earth, $G$ is Newton's gravitational constant, $c$ is the speed of light, $f_{\rm gw}$ is the \gwh frequency, and $\mathcal{M} \equiv (m_1 m_2)^{3/5} /(m_1+m_2)^{1/5}$ is the chirp mass of the binary system ($m_1$ and $m_2$ being the mass of the two components).   
The spin-up $\dot{f}_{\rm gw}$ of this system can be modelled as:
\begin{eqnarray}
    \dot{f}_{\rm gw}&=&\frac{96}{5}\pi^{8/3}\left(\frac{G\mathcal{M}}{c^3}\right)^{5/3} f_{\rm gw}^{11/3}~\label{chirp_pl}  \\
    &\simeq& 9.83\times 10^{-11} \text { Hz/s} \left(\frac{\mathcal{M}}{10^{-6} M_\odot}\right)^{5/3} \left(\frac{f_{\rm gw}}{50 \text{ Hz}}\right)^{11/3}. \nonumber
\end{eqnarray}
These equations will allow us to translate \cwh upper limits, i.e. values of $h_0(f_{\rm gw})$ averaged over the sky at a chosen confidence level, to constraints on the distance reach for PBH inspirals for given chirp masses.

One pipeline, \texttt{Powerflux}, \cite{LIGOScientific:2007hnj,LIGOScientific:2018gpj} performed an all-sky search using data from O3a and looked for \cwh signals from isolated, slowly spinning down or spinning up compact objects (e.g. neutron stars). No signal was found, but upper limits at $95\%$ confidence have been derived across the detector frequency band 20-2000 Hz with a maximum spin-up of $\dot{f}_{\rm max}=1\times 10^{-9}$ Hz/s for an observation time of around six months \cite{LIGOScientific:2021tsm}. Briefly, \texttt{Powerflux} sums the power in many Fourier Transforms after applying Doppler corrections for each template in a large template bank over sky location, frequency, and spin-down/spin-up. When the method does not detect a signal, upper limits are calculated using the maximum strain power over the entire sky in each frequency sub-band.

In addition to constraining the maximum spin-up, the \texttt{Powerflux} pipeline restricts its search to signals whose frequency evolves linearly. For binary black holes, if $f_{\rm gw}$ evolves slowly enough due to a small enough $\mathcal{M}$, we can consider $\dot{f}_{\rm gw}$ to be constant during the observation time, 
and model the frequency evolution as a Taylor series expansion of the integral of equation \ref{chirp_pl} over the observation time:
\begin{equation}
f_{\rm gw,L}(t)=f_0+(t-t_0) \dot{f}_{\rm gw}
\label{eq:linfevol}
\end{equation}
where $f_0$ is the \gwh frequency at a reference time $t_0$ and $f_{\rm gw,L}$ refers to the \gwh frequency calculated assuming a constant spin-up.  Here, we focus on PBH binaries in the linear regime, in which equation \ref{eq:linfevol} is valid.

To obtain the restrictions on spin-up and linearity, we start with the population-averaged $h_0(f)$ values (the green dotted line in figure 4 of \cite{LIGOScientific:2021tsm}), and calculate the corresponding distance and $\dot{f}$ using $h_0$, $f$, and a chosen $\mathcal{M}$ with equations \ref{hamp} and \ref{chirp_pl}, respectively. 
We impose the above-mentioned spin-up condition from the \texttt{Powerflux} search:
\begin{equation}
    \dot{f}\leq 1.00\times 10^{-9} \text{ Hz/s.}
    \label{condspinup}
\end{equation}
We then check that the signal's frequency evolution is linear, based on the spin-up, the observation time $\Tobs$ and the Fast Fourier Transform length $\TFFT$ by ensuring the following condition is true:
\begin{equation}
        \left|f_{\rm gw,L}(\Tobs)- f_{\rm gw}(\Tobs)\right|\leq \Delta f=\frac{1}{\TFFT}
    \label{condlin}
\end{equation}
$\Delta f$ is the frequency bin size, and
$\TFFT=7200$ s for frequencies up to 475 Hz, but could be as low as 1800 s at higher frequencies \cite{LIGOScientific:2021tsm}.
When both conditions given in equations \eqref{condspinup} and \eqref{condlin} are satisfied, we can use that $h_0(f)$ point to derive constraints on the distance reach, and later, on the rates and abundances of PBHs. 

In figure \ref{fig:spinup_linearity}, we can see the effects of imposing these conditions. Note that the upper limits in \cite{LIGOScientific:2021tsm} extend up to 2000 Hz, but only those up to $\sim 250$ Hz are compatible with the conditions of linearity and maximum spin-up imposed by \texttt{Powerflux}. Lower chirp masses imply lower \gwh amplitudes and lower spin-ups, which collectively lead to more constraints at higher frequencies but smaller distance reaches. As chirp mass increases, the maximum spin-up condition becomes more important, further restricting the frequencies at which our analysis can be performed. In effect, the constraint on maximum spin-up and linearity restrict the applicability of this method to systems with $\mathcal{M}\lesssim 10^{-4}M_{\odot}$.

\begin{figure}
    \centering
    \includegraphics[width = \linewidth]{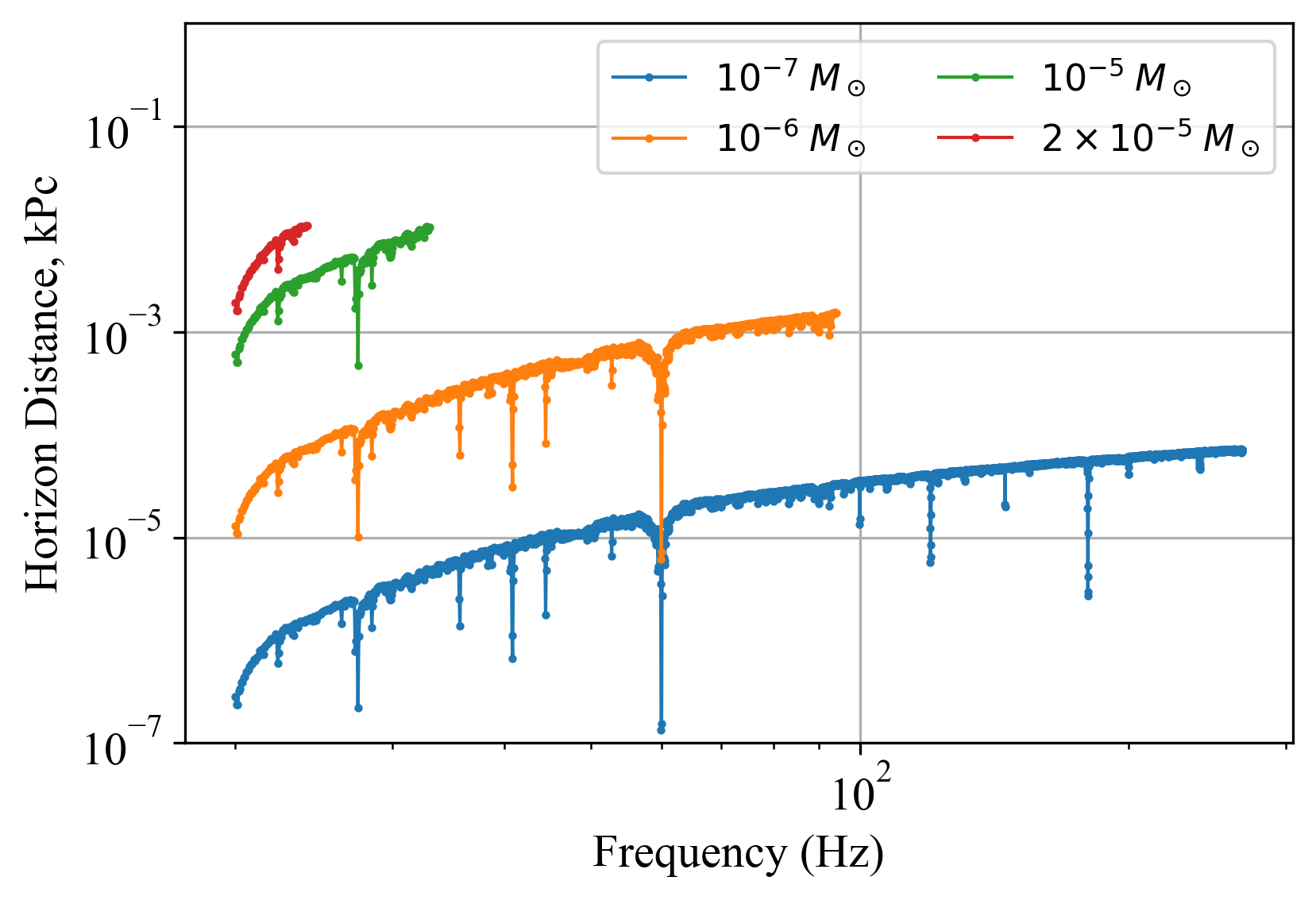}
    \caption{Impact of spin-up condition and linearity condition: for each chirp mass, limits can only be obtained at frequencies that satisfy both equations \ref{condspinup} and \ref{condlin}.}
    \label{fig:spinup_linearity}
\end{figure} 

Using this method, we present the astrophysical reach for planetary-mass inspiraling black holes in Figure \ref{fhdist_constraints}. 
 At lower frequencies ($\sim 20-30$ Hz), the condition on the maximum allowed spin-up permits higher chirp masses, which implies greater accessible distances than those at higher frequencies, e.g. up to 10 parsecs for \(\mathcal{M} = 10^{-5} M_\odot\). At higher frequencies ($\sim 200-250$ hz), the chirp mass of the binary satisfying the above conditions is so small, of $\mathcal{O}(10^{-7})M_\odot$, that in order to achieve the minimum amplitude that was calculated in the \texttt{Powerflux} all-sky search, the binary should have been in the solar system or its vicinity. Nevertheless, in this work we combine the contributions at each frequency to constrain the merger rates of these systems at each chirp mass, as explained in the next section.

\begin{figure}
    \centering
    \includegraphics[width=\columnwidth]{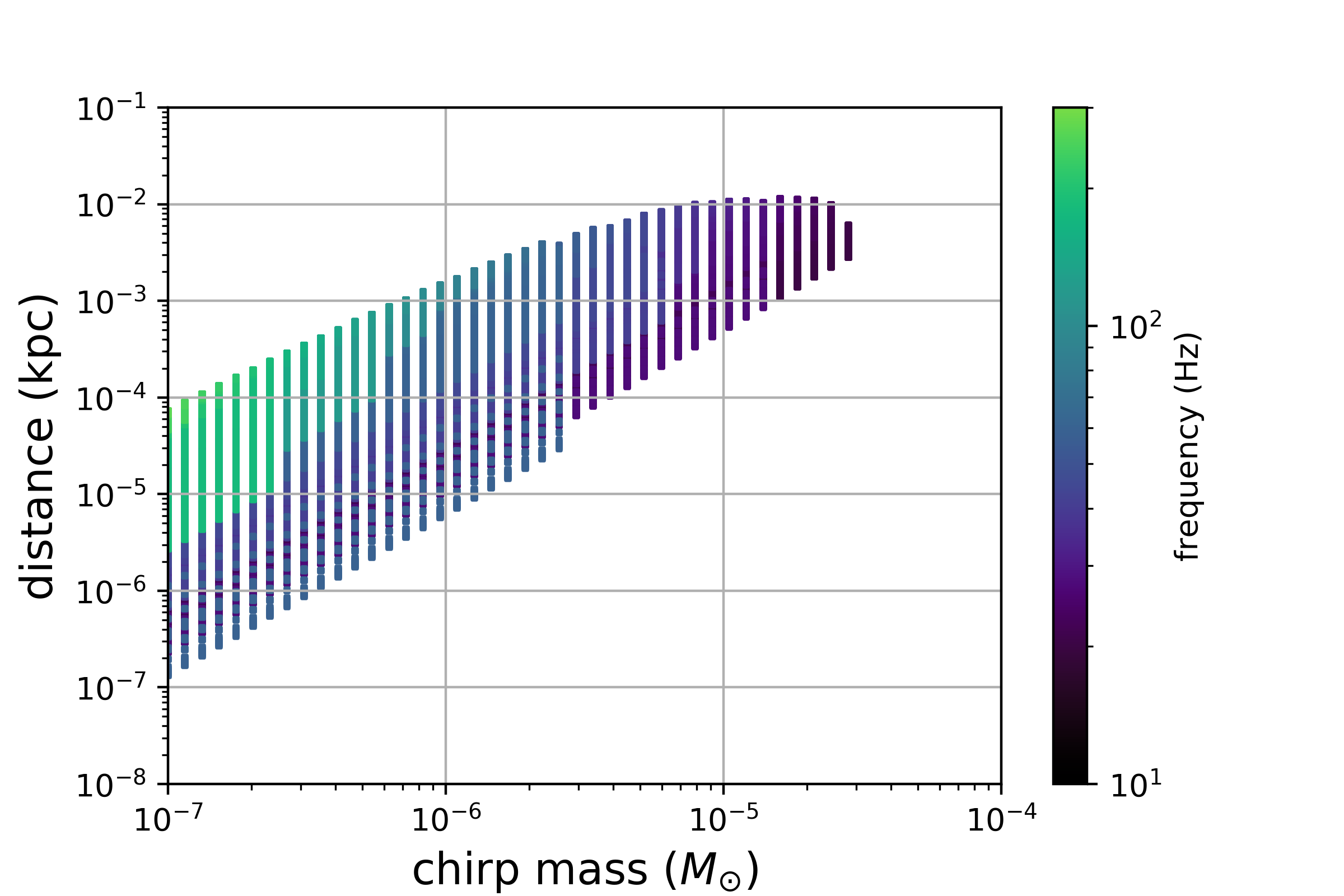}
    \caption{Distance reach as a function of chirp mass with frequency on the color axis, computed using equation \ref{hamp} using as input the upper limits from O3a \cite{LIGOScientific:2021tsm} for chirp-mass-frequency combinations that satisfy $\dot{f}<\dot{f}_{\rm max}$ and evolve linearly in time. The maximum reach is of $\mathcal{O}(\rm pc)$ for most of the planetary-chirp-mass black hole inspirals. Each point contributes to equations \ref{eqn:Nbin} and \ref{eqn:ntot} used to obtain constraints on the rates at a given chirp mass.}
    \label{fhdist_constraints}
\end{figure}

\section{Limits on PBH merging rate and abundance}

{We now estimate upper limits on merger rates as well PBH abundance. }{For a given merger rate \(R\), the number of detectable sources emitting \gws between the frequencies \(f\) and \(f+\delta f\) can be written as:}
\be
N_{\rm bin} (f) \simeq \frac 4 3 \pi d(f)^3 R T,
\label{eqn:Nbin}
\ee
{where \(d(f)\) is the horizon distance, \( T\) is the time period over which the merger events have been integrated, \(T = \mathrm{max}(T_\mathrm{obs},\Delta T)\), \(T_\mathrm{obs}\) is the observation time and \(\Delta T\) is the time spent by the binary system in a given frequency range $\delta f$:} 
\begin{eqnarray}
  \Delta T &=& \frac{5}{256}\pi^{-8/3}\left(\frac{c^3}{G\mathcal{M}}\right)^{5/3} \left[f^{-8/3}-(f+\delta f)^{-8/3 }\right].~ 
\label{eqn:deltaT}
\end{eqnarray}
In cases where $\Delta T$ exceeds $\Tobs$, the number of \cwh sources probed by the detector is largely dominated by the number of sources emitting in the frequency band  $(f,f+\delta f)$, including those that started emitting much before the start of the survey. Here, $\delta f=\dot{f}_{\rm max}\Tobs=0.01$ Hz and represents the maximum linear frequency shift that the binary system could have during O3a.

{Given $N_{\rm bin} (f)$, the total number of binaries at a given chirp mass is then calculated by summing the number of binaries at each frequency $f_i$:}

\be
N_{\rm bin}^{\rm tot} = \sum_{ i} N_{\rm bin} (f_i)~.
\label{eqn:ntot}
\ee
{In the absence of observations, the above expression can be inverted to obtain an upper limit on merger rate given the calculated astrophysical reach.}

Figure \ref{fig:rates} shows upper limits on merging rates in the solar system vicinity, on the basis of all-sky \cw upper limits in O3a.  Though these limits are not stringent due to the low distance reach of our detectors to these systems, they show that \cwh observations can be mapped to constraints on the merging rates, which tend to be dominated by early binaries in the planetary-mass range.
Furthermore, by using the cosmological rate predictions for early PBH binaries and of PBH binaries in clusters, these limits can be translated into constraints on the dark matter fraction of PBHs. 
We use the analytical prescriptions from~\cite{raidal2019formation,Hutsi:2020sol} for the cosmological merger rates that assume a purely Poissonian PBH spatial separation at formation, given by:
\bea
        R^{\rm cos}_{\rm prim}  &\approx & \frac{1.6 \times 10^6}{\rm Gpc^3 yr} f_{\rm sup} f_{\rm PBH}^{{53/37}} \left(\frac{m_1 + m_2}{M_\odot}\right)^{-32/37} \nonumber \\ & \times & \left[\frac{m_1 m_2}{(m_1+m_2)^2}\right]^{-34/37} f(m_1) f(m_2)~,  \label{eq:cosmomerg}
\eea
which correspond to the rate per unit of logarithmic mass of the two binary black hole components $m_1$ and $m_2$. $f_{\rm PBH}$ is the dark matter density fraction made of PBHs and $f(m)$ is the density distribution of PBHs normalized to one ($\int f(m) {\rm d} \ln m = 1$).  We have included a suppression factor $f_{\rm sup}$ that effectively takes into account a rate suppression due to the gravitational influence of early forming PBH clusters, nearby PBHs and matter inhomogeneities~\cite{raidal2019formation}. Finally, the expected merging rates near the solar system are calculated by assuming a constant local dark matter density of $\sim 10^{16} M_\odot {\rm Mpc}^{-3}$ \cite{Weber:2009pt} that is consistent with the galactic dark matter profile. In this way, we obtain that $R = 3.3 \times 10^5 R_{\rm prim}^{\rm cos}$~\cite{Miller:2020kmv}.

For equal-mass PBHs and $f_{\rm PBH}=1$, one may expect $f_{\rm sup} \approx 2 \times 10^{-3}$ \cite{Hutsi:2020sol,Clesse:2020ghq,raidal2019formation} but the exact value of $f_{\rm sup}$ for wide mass functions and binaries with high mass ratios is largely uncertain and model dependent. For example, external tidal fields and highly eccentric orbits of low-mass PBHs may affect this parameter {\cite{Eroshenko:2016hmn,Cholis:2016kqi}}. However, due to the heavy model-dependence of \(f_{\rm sup}\), we provide limits on an effective parameter $\tilde{f}$ that is defined to be model-agnostic by including \(f_{\rm sup}\). \begin{equation}
\tilde f^{53/37} \equiv f_{\rm sup} f(m_1) f(m_2) f_{\rm PBH}^{53/37}.\label{ftilde}
\end{equation}

First, we consider equal-mass PBH mergers, for which 
\begin{eqnarray}
R =& 1.04 \times 10^{-6}\, \mathrm{kpc}^{-3} \mathrm{yr}^{-1} f_{\rm sup} f(m_{\rm PBH})^2 \nonumber\\
&\left(\frac{m_\mathrm{PBH}}{M_\odot}\right)^{-32/37}  \left(f_{\rm PBH}\right)^{53/37}.
\label{eqn:rate}
\end{eqnarray}
The upper limits on $\tilde{f}$  obtained by considering equal mass binaries are shown in Figure~\ref{fig:rates}. Since $f_{\rm PBH}$ is by definition less than 1, the upper limit set by this search is trivially satisfied. We therefore conclude that \cw methods cannot yet probe equal-mass PBH binaries.   

This search can also be used to place constraints on binary systems with highly asymmetric mass ratios for the same chirp masses, for which the expected merger rates are higher by several orders of magnitude. Given that an important motivation for PBHs come from observations of compact binary coalescences in the stellar-mass range, we consider systems with $m_1 = 2.5\, M_\odot$, as motivated by the QCD transition~\cite{Byrnes:2018clq,Carr:2019kxo,Clesse:2020ghq} and some mergers such as GW190425 and GW190814~\cite{Clesse:2020ghq}, in a binary with a much lighter PBH of mass $m_2$, which could be probed with \cws. We therefore consider the merging rate in the limit $m_1 \gg m_2$:
\begin{eqnarray}
R =& 5.28 \times 10^{-7}\, \mathrm{kpc}^{-3} \mathrm{yr}^{-1} f_{\rm sup} f(m_1) f(m_2) \nonumber\\
&\left(\frac{m_1}{M_\odot}\right)^{-32/37} \left(\frac{m_2}{m_1}\right)^{-34/37} \left(f_{\rm PBH}\right)^{53/37}~.
\label{eqn:rate_asymm}
\end{eqnarray}
The limits on \(\tilde f\) for asymmetric mass ratios is shown in Figure \ref{fig:rates_asymm} as a function of \(m_2\). 
For asymmetric PBH mass ratios, if we assume that $f(m_1) \approx 1$ at $2.5 M_\odot$, as is expected for broad mass functions imprinted by the QCD transition, then $\tilde f = f_{\rm PBH} f(m_2)^{37/53} f_{\rm sup}$.  In a model where $f_{\rm PBH} =1$, these limits can be translated into the probability density function of PBHs of mass $m_2$, given by $f(m_2)$ in our benchmark scenario for rate suppression using the above equations. The upper limits on $f(m_2)$ are also shown in Figure \ref{fig:rates_asymm} on the color axis. Given the normalization of \(f(m)\), these upper limits are trivially satisfied. {We find that $\tilde f \lesssim \mathcal O(10)$ in the range $10^{-10} M_\odot < m_2 < 10^{-8} M_\odot$, does not yet allow us to probe $f_{\rm PBH} \lesssim 1$, though it is not far from setting an interesting limit in the asteroid-mass range ($m_2 \lesssim 10^{-11} M_\odot$) that remains unconstrained by other probes.
It will therefore be relevant to pursue continuous-wave searches to constrain PBHs with LIGO/Virgo or with the third generation of ground-based detectors, such as Einstein Telescope or Cosmic Explorer, {in which we expect around an order of magnitude improvement in our estimation of $\tilde f$ relative to LIGO/Virgo design sensitivity \cite{Maggiore:2019uih}. Such an increase in sensitivity would imply that $\tilde f$ would dip below one, that our distance reach and detectable volume of space would increase by one and three orders of magnitude, respectively, compared to the reach we derive here.}

{Future constraints would allow us to either confirm or exclude models that allow for minimal rate suppression, i.e. $f_{\rm sup}\gtrsim\mathcal{O}(0.1)$. One such model, that accounts for both disruptions due to matter fluctuations and nearby PBHs at distances
smaller than the maximal comoving distance that would fall onto the binary before matter-radiation equality, predicts a range of suppression factors between [0.2,1] for $f_{\rm PBH}>0.1$ \cite{Hutsi:2020sol,Phukon:2021cus}. Additional models predict minimal work and torques on PBH binaries due to the tidal field viralized halos and to interactions with surrounding cold dark matter \cite{Ali-Haimoud:2017rtz,Kavanagh:2018ggo}, implying negligible rate suppression. }

\begin{figure}
    \centering
    \includegraphics[width=\columnwidth]{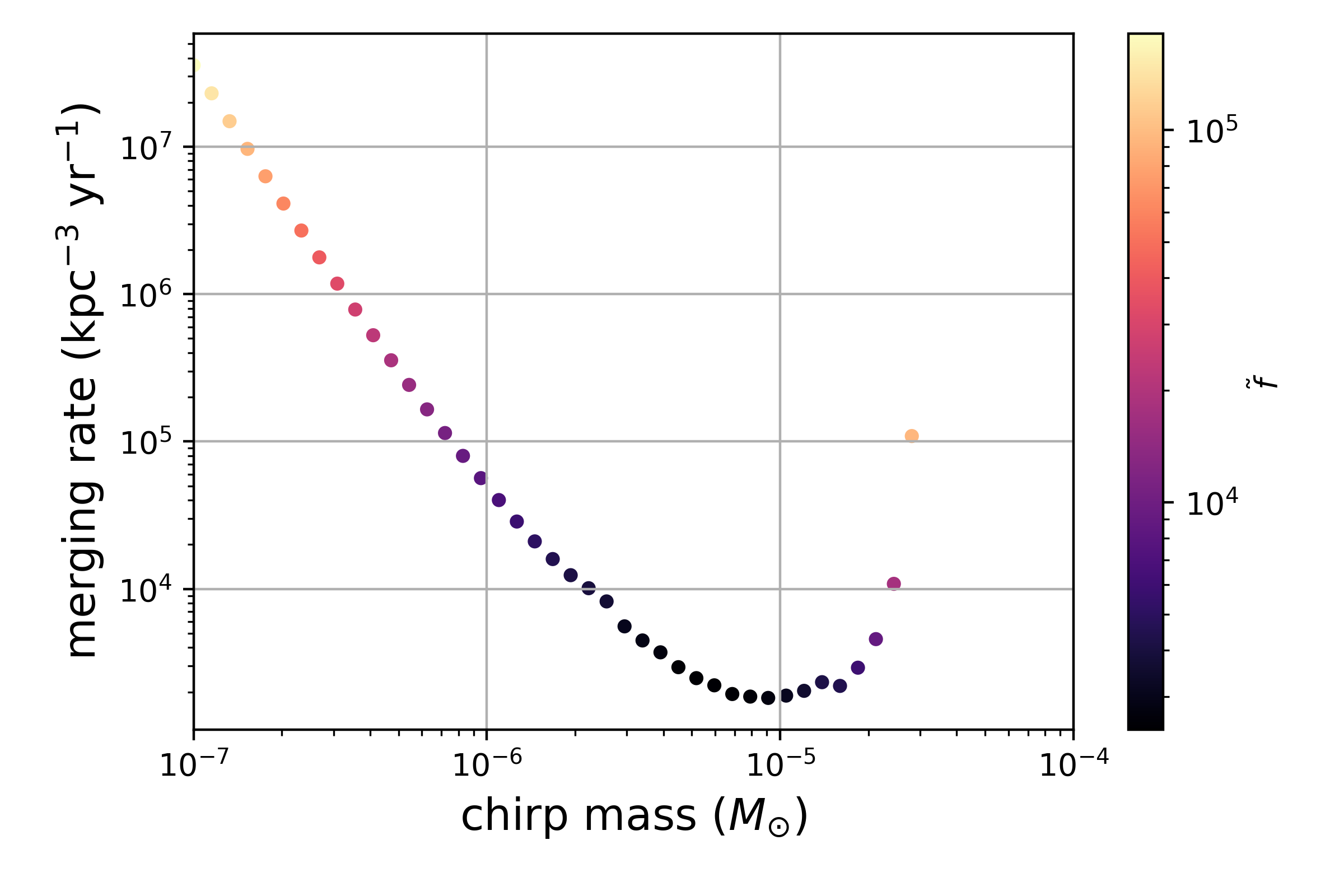}
    \caption{Upper limit on merging rates of planetary and asteroid chirp mass binaries. These rates do not depend on any particular PBH formation model; they only depend on the sensitivity of the \cwh search, i.e. the distance reach. Also shown in color are upper limits on the fraction of dark matter that primordial black holes could compose assuming rate models for equal-mass primordial black hole binaries.}
    \label{fig:rates}
\end{figure}

\begin{figure}
    \centering
    \includegraphics[width=\columnwidth]{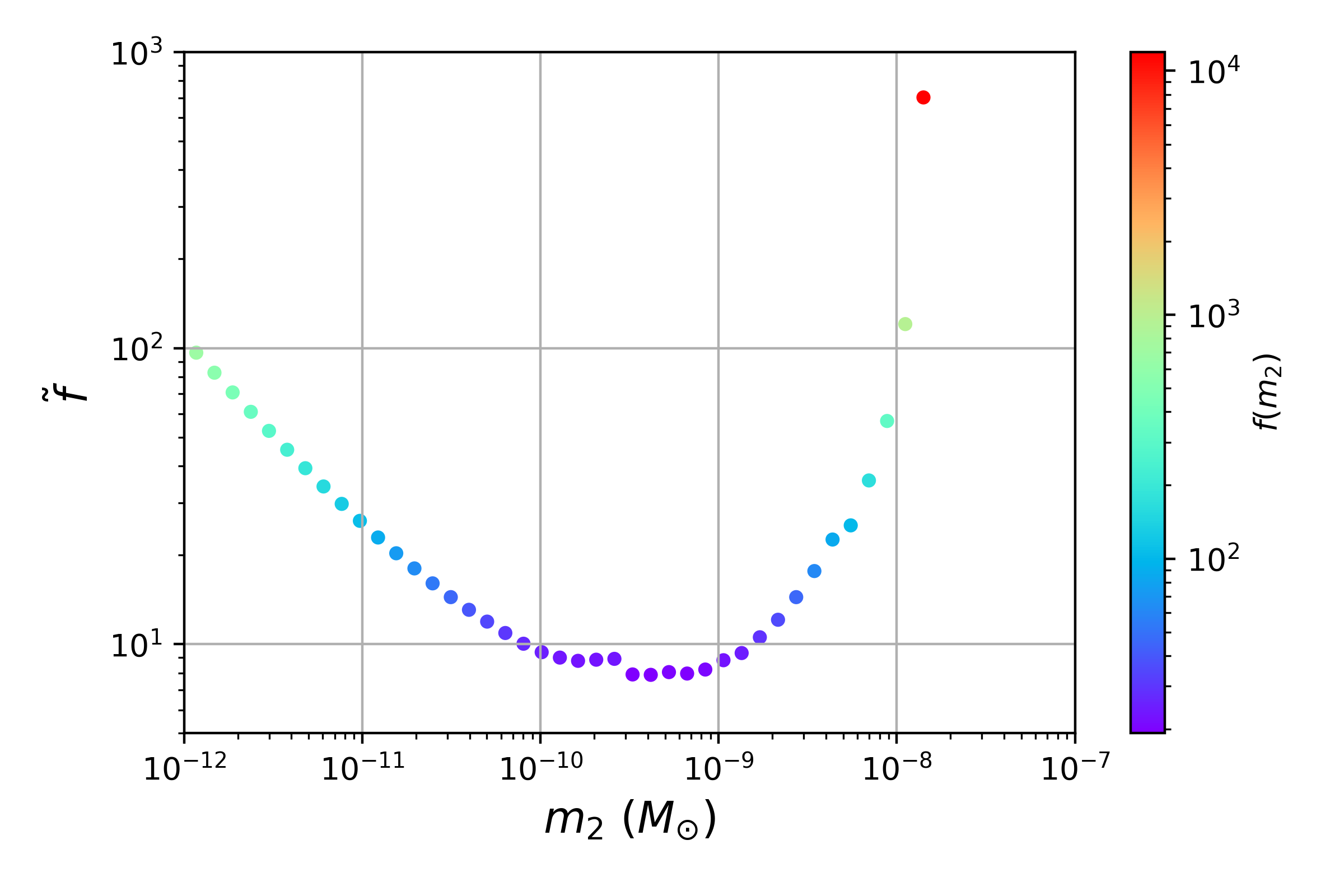}
    \caption{Upper limit on the PBH abundance for asymmetric mass ratio binaries with one component at \(m_1 = 2.5 M_\odot\). Also shown in color is upper limit on the mass function of the smaller companion PBH assuming \(f(m_1) = 1\) and \(f_{\rm PBH} =1\).}
    \label{fig:rates_asymm}
\end{figure}

\section{Discussion}
For the first time, upper limits on \cwh emission from isolated neutron stars have been mapped to constraints on planetary-mass and asteroid-mass black holes that could be of primordial origin. Though our results are valid at only nearby distances, they highlight the ease of adapting \cwh and transient \cwh searches to constrain the presence of primordial black holes. Our work thus motivates both the extension of \cwh searches to higher spin-ups, e.g. \cite{LIGOScientific:2016bah}, which would allow more stringent constraints at higher frequencies, and the need to develop other ways to probe primordial black holes in and above this mass range, {especially the asteroid-mass range, which is not well constrained \cite{Katz:2018zrn,Montero-Camacho:2019jte}}. Based on the work in Ref. \cite{Miller:2020vsl}, the parameter space and astrophysical distance reach covered by transient \cwh searches are larger than those from standard \cwh ones. The techniques presented in this paper, coupled with transient \cwh search methods, could provide interesting constraints on primordial black holes without having to develop an entirely new infrastructure to run such searches. 

Looking towards the future, we note that signals from stellar-mass binary black holes will be detectable by future detectors \cite{wyithe2003low,gair2011exploring}, such as LISA \cite{danzmann2003lisa}, Einstein Telescope \cite{punturo2010third}, DECIGO \cite{Kawamura:2020pcg} and TianQin \cite{Luo:2015ght}. The mapping performed here, as well as the \cwh and transient \cwh methods discussed, will be useful to detect these kinds of sources, {or provide upper limits on asteroid-mass PBHs whose contributions to dark matter are not constrained}.

\section*{Acknowledgements} This research has used data obtained from the
Gravitational Wave Open Science Center, a service of
LIGO Laboratory, the LIGO Scientific Collaboration and
the Virgo Collaboration. LIGO is funded by the U.S.
National Science Foundation. Virgo is funded by the
French Centre National de Recherche Scientifique
(CNRS), the Italian Istituto Nazionale della Fisica
Nucleare (INFN), and the Dutch Nikhef, with contributions
by Polish and Hungarian institutes. 

We thank the LIGO/
Virgo Collaboration continuous-wave working group, the Rome Virgo group, and Bernard Whiting for
helpful discussions.

Computational resources have been provided by the supercomputing facilities of the Université catholique de Louvain (CISM/UCL) and the Consortium des Équipements de Calcul Intensif en Fédération Wallonie Bruxelles (CÉCI) funded by the Fond de la Recherche Scientifique de Belgique (F.R.S.-FNRS) under convention 2.5020.11 and by the Walloon Region.

This material is based upon work supported by NSF’s LIGO Laboratory which is a major facility fully funded by the National Science Foundation.

A.L.M. is a beneficiary of a FSR Incoming Postdoctoral Fellowship. F.D.L. is supported by a FRIA Grant of the Belgian Fund for Research,  F.R.S.-FNRS. {NA is supported by the NSF grant PHY-1806671 and a CIERA Postdoctoral Fellowship from the Center for Interdisciplinary Exploration and Research in Astrophysics at Northwestern University. }



\bibliographystyle{ieeetr}
\bibliography{biblio}

\end{document}